\documentclass[amsmath,amssymb,twocolumn,nofootinbib]{revtex4}

\usepackage{latexsym,comment,verbatim}
\usepackage{amssymb,multirow}
\usepackage{amsfonts}
\usepackage{amsmath,color}
\usepackage{graphicx}
\usepackage{bm}

\def\mb#1{\mathbf{#1}}

\def\ber{\begin{eqnarray}}
\def\eer{\end{eqnarray}}
\def\beq{\begin{equation}}
\def\eeq{\end{equation}}

\def\rmd{{\rm d}}

\def\ed{\end{document}}

\def\dTT#1{\frac{\mathrm{d} ^{2}#1}{\mathrm{d}T^{2}}}

\def\dtau#1{\frac{\mathrm{d} #1}{\mathrm{d}\tau}}
\def\dttau#1{\frac{\mathrm{d} ^{2}#1}{\mathrm{d}\tau^{2}}}
\def\sT{\sin \left(\omega T \right)}
\def\cT{\cos \left(\omega T \right)}

\def\bfbeta{\pmb{\beta}}
   \let\square=\dal
\newcommand{\ppar}[2]{\frac{\partial #1}{\partial #2}}

\begin{document}

\author{Matteo Luca Ruggiero}
\email{matteo.ruggiero@polito.it}
\affiliation{Politecnico di Torino, Corso Duca degli Abruzzi 24, 10129 Torino - Italy \\ INFN - LNL , Viale dell'Universit\`a 2, 35020 Legnaro (PD), Italy}

\date{\today}

\title{A note on the gravitoelectromagnetic analogy}

\begin{abstract}
We discuss the {linear} gravitoelectromagnetic approach used to solve Einstein equations in the weak-field and slow-motion approximation, which is a powerful tool to explain, by  analogy with electromagnetism, several gravitational effects in the Solar System, where the approximation holds true. In particular, {we discuss the analogy according to which Einstein equations can be written as Maxwell-like equations  and focus on the definition of the gravitoelectromagnetic fields in non stationary conditions.  Furthermore, we examine to what extent, starting from a given solution of Einstein equations, gravitoelectromagnetic fields can be used to describe  the motion of test particles using a Lorentz-like force equation} 
\end{abstract}

\maketitle

\section{Introduction}\label{sec:intro}

Einstein's theory of gravitation, General Relativity (GR), completely changed our view and understanding of space and time and of the
interplay between them. For these reasons, soon  after its publication, GR deeply influenced  scientific and philosophical thought, even if only few and non-highly accurate observational evidences were available. As emphasised by \citet{will2018theory}, several events, pertaining to both the development of the theoretical framework and the observations, contributed to establish the basis of experimental gravitation, starting from the beginning of the '60s. At first,  the great majority of the experimental tests were performed within the Solar System; subsequently, observations involving sources outside the Solar System were available: in the latter case, we often deal with extreme events producing huge perturbations in the fabric of space-time. On the contrary, in the Solar System, the gravitational field is weak but, nonetheless, GR successfully predicts the existence of new phenomena for which Newtonian gravity is inadequate.  

Einstein equations in the Solar System can be adequately solved in  weak-field approximation (small masses, low velocities); in particular, these equations can  be written in analogy with Maxwell equations for the electromagnetic fields, where the mass density and current play the role of the charge density and current, respectively \cite{Ruggiero:2002hz,Mashhoon:2003ax}.  As a consequence, a \textit{gravitomagnetic} field arises, due to mass currents; more in general, every theory that combines Newtonian gravity with Lorentz invariance predicts the existence of these gravitomagnetic effects. Interestingly enough, the existence of a magnetic-like part of the gravitational field was already suggested by Heaviside, at the end of 1800,  on the basis of the similarity between Newton's law of gravitation and Coulomb's law of electrostatic force (see \citet{McDonald:1997fd} and references therein).  This analogy can be exploited to explain GR effects in terms of electromagnetic ones:  this is the case, for instance,   of the famous Lense-Thirring gyroscope precession \cite{Iorio:2010rk}, which can be explained in analogy with the precession of a magnetic dipole in a magnetic field. 

However, we must always remember that GR is a  non linear theory, so the use of the {linear} gravitoelectromagnetic (GEM) analogy has some limitations which need to be emphasised.  To this end, it is useful to remember that it is also possible to develop an exact gravitoelectromagnetic analogy in full GR  {(see e.g. \citet{cattaneo1958general,costa2008gravitoelectromagnetic,Mashhoon:1996wa,Ramos:2006ho,Costa:2012cw,chicone2002generalized,RR2004,jantzen1992many,lynden1998classical}} and also the recent publication by \citet{Costa:2021atq}).  
The purpose of this paper is to discuss in full details the linear GEM analogy and its limitations. In particular, in Section \ref{sec:eqGEM} we discuss in some details the customary approach which, starting from Einstein equations in weak-field and slow-motion approximation, leads to the definition of the gravitoelectromagnetic fields. In Section \ref{sec:motion} we consider the geodesic equation, for a given solution of Einstein equations {and discuss under which hypotheses it can be formally expressed in terms of a Lorentz-like force equation for test masses; then} we focus on an application of this formalism to the spacetime of a plane gravitational wave. Discussion and conclusions are {given} eventually in Section \ref{sec:disconc}.

\section{{Linear } gravitoelectromagnetic form of Einstein equations}\label{sec:eqGEM}

Let us start from  Einstein equations
\beq
G_{\mu\nu}=\frac{8\pi G}{c^{4}}T_{\mu\nu}. \label{eq:Einstein0}
\eeq
In the weak-field approximation, the gravitational field can be considered a perturbation of flat spacetime, described by the Minkowski tensor $\eta_{\mu\nu}$.\footnote{The spacetime signature is $(-1,1,1,1)$; Greek indices run from 0 to 3, while Latin indices from 1 to 3; boldface symbols like $\mb x$ refer to space vectors.}
As a consequence, the metric tensor can be written in the form $g_{\mu\nu}=\eta_{\mu\nu}+h_{\mu\nu}$, where $h_{\mu\nu}$ is a weak perturbation:  {$|h_{\mu\nu}|\ll 1$}. If we introduce $\bar h_{\mu\nu}=h_{\mu\nu}-\frac 1 2 \eta_{\mu\nu}h$, where $h=h_{\mu}^{\ \mu}$, Einstein equations (\ref{eq:Einstein0}) become (see e.g. \citet{straumann2013applications})
\beq
-\square{ \bar h}_{\mu\nu}-\eta_{\mu\nu}\bar h_{\alpha\beta}^{\ \ \ ,\alpha\beta}+\bar h_{\mu\alpha, \nu}^{\ \ \ \ \ \alpha}+\bar h_{\nu\alpha, \mu}^{\ \ \ \ \ \alpha}=\frac{16\pi G}{c^{4}}T_{\mu\nu}.  \label{eq:Einstein1}
\eeq
The gauge freedom can be exploited setting   the \textit{Hilbert gauge condition}
\beq
\bar h^{\mu\nu}_{\ \ ,\nu}=0. \label{eq:Hilibert}
\eeq
{The above condition is also known as \textit{Einstein gauge, de Donder gauge, Fock gauge, or Lorentz gauge\cite{Carroll:1997ar}};  in particular, the latter name refers to the analogy with the correspondent condition used in electromagnetism (see below)}. Then,  from (\ref{eq:Einstein1}) we get
\beq
\square{ \bar h}_{\mu\nu}=-\frac{16\pi G}{c^{4}}T_{\mu\nu},  \label{eq:Einsteinweak}
\eeq
Notice that the  condition (\ref{eq:Hilibert}) can be always achieved by a gauge transformation; in fact, Einstein equations are invariant with respect to the infinitesimal transformations
\beq
h_{\mu\nu} \rightarrow h_{\mu\nu}+\xi_{\mu,\nu}+\xi_{\nu,\mu} \label{eq:gauge1}
\eeq
which, in terms of $\bar h_{\mu\nu}$ becomes
\beq
\bar h_{\mu\nu} \rightarrow \bar h_{\mu\nu}+\xi_{\mu,\nu}+\xi_{\nu,\mu}-\eta_{\mu\nu}\xi^{\alpha}_{\ ,\alpha} \label{eq:gauge2}
\eeq
So, if $\bar h^{\mu\nu}_{\ \ ,\nu}\neq0$, it is sufficient to choose $\xi^{\mu}$ to be a solution of $\square \xi^{\mu}=-\bar h^{\mu\nu}_{ \ \ ,\nu}$.

Eqs. (\ref{eq:Einsteinweak})  are in clear analogy with Maxwell equations for the electromagnetic four-potential: so, they can be solved in the same way  {(see e.g. \citet{Ruggiero:2002hz,Mashhoon:2003ax,Mashhoon:2001ir,Mashhoonspin}, \citet{padmanabhan2010gravitation})}. In fact, neglecting the solution of the homogeneous wave equations associated to (\ref{eq:Einsteinweak}),  the general solution is given in terms of retarded potentials
\beq
 {\bar h}_{\mu\nu}=\frac{4G}{c^4}\int_{V} \frac{T_{\mu\nu}(ct-|{\mathbf x}-{\mathbf x}'|, {\mathbf x}')}{|{\mathbf x}-{\mathbf x}'|}\rmd^3 x'\ , \label{eq:solgem1}
\eeq
where integration is extended to the volume $V$, containing the source.  We may set $T^{00}=\rho_{  } c^2$ and $T^{0i}=cj_{  }^i$, in terms of the mass density $\rho_{  }$ and mass current $j_{  }^{i}$ of the source, so that   $j_{  }^\mu=\left(c\rho_{  },j_{g}^{i} \right)=\left(c\rho_{  },{\mathbf j_{  }}\right)$ is the mass-current four vector  of the source. Since, in linear approximation, $\displaystyle T^{\mu\nu}_{\ \ ,\nu}=0$, we obtain the continuity equation 
\beq
\ppar{\rho}{t}+\bm \nabla \cdot \mb j=0 \label{eq:continuity}
\eeq
If we assume that the source consists of a finite distribution of slowly moving matter, with $|\mb v|\ll c$, then $T_{ij} \simeq \rho v_{i}v_{j}+p\delta_{ij}$, where $p$ is the pressure: from (\ref{eq:solgem1}) we see that $\bar h_{ij} =O(c^{-4})$:  in the linear GEM approach, we neglect in the metric tensor terms that are $O(c^{-4})$.

Consequently, from (\ref{eq:solgem1}) we get
\beq
{\bar h}_{00}=\frac{4G}{c^{2}}\int_{V} \frac{\rho(ct-|{\mathbf x}-{\mathbf x}'|, {\mathbf x}')}{|{\mathbf x}-{\mathbf x}'|}\rmd^3 x'\ , \label{eq:solgemh00}
\eeq
\beq
{\bar h}_{0i}=-\frac{4G}{c^{3}}\int_{V} \frac{j^{i}(ct-|{\mathbf x}-{\mathbf x}'|, {\mathbf x}')}{|{\mathbf x}-{\mathbf x}'|}\rmd^3 x'\ . \label{eq:solgemh0i}
\eeq
The other components of $\bar h_{\mu\nu}$ are zero at the given approximation level.

In analogy with the corresponding solutions of electromagnetism, it is possible to  introduce  the \textit{gravitoelectromagnetic potentials}: namely, the gravitoelectric  $\Phi_{  }$  and gravitomagnetic $A^{i}_{  }$ {potentials} are defined by
\beq
\bar h_{00} \doteq 4\frac{\Phi_{  }}{c^{2}}, \quad \bar h_{0i}=-2 \frac{A_{i}}{c^{2}}, \label{eq:defphiAigem}
\eeq
which, taking into account Eqs. (\ref{eq:solgemh00}) and (\ref{eq:solgemh0i}), take the form
\beq
\Phi ={G}\int_{V} \frac{\rho(ct-|{\mathbf x}-{\mathbf x}'|, {\mathbf x}')}{|{\mathbf x}-{\mathbf x}'|}\rmd^3 x'\ , \label{eq:solgemphi1}
\eeq
\beq
A_{i}=\frac{2G}{c}\int_{V} \frac{j^{i}(ct-|{\mathbf x}-{\mathbf x}'|, {\mathbf x}')}{|{\mathbf x}-{\mathbf x}'|}\rmd^3 x'\ . \label{eq:solgemAi1}
\eeq
Eventually, the spacetime metric describing the solutions of Einstein's equation in weak-field approximation is written in the form \cite{Ruggiero:2002hz,Mashhoon:2003ax}
\beq
\mathrm{d} s^2= -c^2 \left(1-2\frac{\Phi}{c^2}\right)\rmd t^2 -\frac4c A_{i}\rmd x^{i}\rmd t +
 \left(1+2\frac{\Phi}{c^2}\right)\delta_{ij}\rmd x^i \rmd x^j. \  \label{eq:weakfieldmetric1}
\eeq

Now that we have defined the gravitoelectromagnetic potentials, it is possible to reconsider the Hilbert gauge condition (\ref{eq:Hilibert}) and express it in terms of $\Phi_{  }$ and $A^{i}_{  }$.  From (\ref{eq:Hilibert}) we obtain indeed two conditions: setting $\mu=0$ we get
\beq
\bar h^{00}_{\ \ ,0}+ \bar h^{0i}_{\ \ ,i}=0 \rightarrow  \frac 1 c \frac{\partial \Phi}{\partial t}+\frac 1 2 \bm \nabla \cdot \mb A=0, \label{eq:gauge11}
\eeq
which is the same as the {Lorenz gauge} condition for electromagnetic fields. {If we consider the space part  ($\mu=i$) of the gauge condition (\ref{eq:Hilibert}), we obtain 
\beq
\bar h^{i0}_{\ \ ,0}+\bar h^{ij}_{\ \ ,j}=0 \rightarrow  \frac{2}{c^{3}}\frac{\partial A^{i}}{\partial t}+\bar h^{ij}_{\ \ ,j}=0.
\label{eq:gauge120}
\eeq
 {Indeed, even if the terms $\bar h_{ij}$ are not displayed in the metric (\ref{eq:weakfieldmetric1}) for being $O(c^{-4})$, they are not necessarily exactly zero: as a consequence, Eq. (\ref{eq:gauge120}) does not imply a time-independent gravitomagnetic potential $\mb A$.\\
\indent The issue of the time-independence of the gravitomagnetic potential has been discussed in several papers in the past, still with no general agreement. For instance \citet{bakopoulos2014gem} explicitly consider $\bar h_{ij}=0$, hence from (\ref{eq:gauge120}) they deduce $\displaystyle \frac{\partial \mb A}{\partial t}=0$, while  \citet{harris1991analogy} maintains that  $\displaystyle \frac{\partial \mb A}{\partial t}=O(c^{-2})$; similar conclusions about the time independence of the gravitomagnetic field are obtained by \citet{clark2000gauge}. For further insights on this topic we refer to the papers by \citet{costa2008gravitoelectromagnetic} and \citet{PascualSanchez:2000rd}.}\\ 
According to the approach used by  \citet{Mashhoon:2003ax,Mashhoon:2001ir,Mashhoonspin}, \citet{Ruggiero:2002hz}, the gravitoelectric $\mb E$ and gravitomagnetic $\mb B$ fields are defined by
\beq
\mb E=-\bm \nabla \Phi-\frac{1}{2c}\frac{\partial \mb A}{\partial t}, \quad  \mb B= \bm \nabla \wedge \mb A, \label{eq:solgemEB1}
\eeq
and both fields can be time-dependent. In addition, taking account the Einstein equations (\ref{eq:Einsteinweak}), we may write the equations for the gravitoelectromagnetic fields in the form
\begin{eqnarray}
\bm \nabla \cdot \mb E&=& 4\pi G \rho, \label{eq:gemm11} \\
\bm \nabla \wedge \mb E&=&-\frac{1}{c}\frac{\partial}{\partial t} \left(\frac{\mb B}{2} \right), \label{eq:gemm21} \\
\bm \nabla \cdot  \left(\frac{\mb B}{2} \right)&=&0, \label{eq:gemm31} \\
\bm \nabla \wedge  \left(\frac{\mb B}{2} \right)&=&\frac{4\pi G}{c} \mb j+ \frac 1 c \ppar{\mb E}{t}. \label{eq:gemm41}
\end{eqnarray}
In particular, from Eqs. (\ref{eq:gemm11}) and (\ref{eq:gemm41}) the continuity equation (\ref{eq:continuity}) is obtained.  We notice the factor $\frac 1 2$ near the gravitomagnetic field $\mb B$, with respect to the original Maxwell equations for the electromagnetic fields: this is due to the tensorial character of the gravitational field in GR  (see \citet{Mashhoon:2001ir})}. 

{It is interesting to point out that if we apply a different gauge condition we obtain different equations for the gravitoelectric and gravitomagnetic field, as discussed for instance by \citet{Bertschinger:1993xt,Costa:2012cw,damour1991general,carroll2019spacetime}.}

{We want to  emphasise here an important point: the definition (\ref{eq:solgemEB1}) of the gravitoelectric field  does not agree with the corresponding one
\beq
\mb E= -\bm \nabla \Phi-\frac{2}{c} \ppar{\mb A}{t}, \label{eq:defEtime0}
\eeq
that we are going to obtain in Section \ref{sec:motion},  writing the geodesic equation in weak-field and slow-motion approximation.
 {Actually, if we use the definition (\ref{eq:defEtime0}), the sources equations for the gravitoelectromagnetic fields get modified: as emphasized by \citet{Costa:2012cw} it is not possible to obtain a one-to-one gravitoelectromagnetic analogy both for the geodesic equation and for the field equations, since in any case non-Maxwellian terms appear. Using the definition (\ref{eq:defEtime0}), a different form of the induction law (\ref{eq:gemm21}) is obtained, which is the same as the one obtained by  \citet{bini2008gravitational}  starting from the gravitoelectromagnetic force  acting on a test particle (see next Section). }

\section{Gravitoelectromagnetic description of the motion of test masses}\label{sec:motion}

Let us suppose that the spacetime metric is written in the quite general form 
\beq
ds^{2}=g_{00}c^{2}dt^{2}+2g_{0i}cdtdx^{i}+g_{ij}dx^{i}dx^{j}. \label{eq:metricastazionaria}
\eeq
By setting
\[
\frac{\Phi}{c^{2}}=\frac{g_{00}+1}{2} \quad \frac{\Psi}{c^{2}}=\frac{g_{ij}-1}{2} \quad \frac{A_{i}}{c^{2}}=-\frac{g_{0i}}{2}
\]
where $|\frac{\Phi}{c^{2}}| \ll 1$, $|\frac{\Psi}{c^{2}}| \ll 1$, $|\frac{A_{i}}{c^{2}}| \ll 1$, the above metric can be written in the form
\beq
\mathrm{d} s^2= -c^2 \left(1-2\frac{\Phi}{c^2}\right)\rmd t^2 -\frac4c A_{i}\rmd x^{i}\rmd t  +
 \left(1+2\frac{\Psi}{c^2}\right)\delta_{ij}\rmd x^i \rmd x^j\ , \label{eq:weakfieldmetric11}
\eeq
We do not require that the starting metric (\ref{eq:metricastazionaria}) has been obtained solving Einstein equations in weak-field approximation. In other words, we assume that a given solution of the field equations can be written in this form, and it represents a small perturbation of flat spacetime.  In particular, the  gravitoelectromagnetic potentials can be time dependent. The relation between $\Phi_{  }$ and $A^{i}_{  }$ and the sources can be of course obtained writing the field equations (\ref{eq:Einstein1}). For instance,  this approach was used by \citet{bini2008gravitational}:   the authors start from a spacetime in the form (\ref{eq:weakfieldmetric11}), assume that the gravitomagnetic potential describes the field of a source whose angular momentum changes with time, and calculate effective sources for the spacetime metric.

{Let us start from the  line element (\ref{eq:weakfieldmetric11}) and calculate the geodesic equation up to linear order in $\bfbeta={\mathbf v}/c$. From
\beq
\dttau x^{\mu}+\Gamma^{\mu}_{\alpha\beta} \dtau{x^{\alpha}} \dtau{x^{\beta}}=0,  \label{eq:geott1}
\eeq
we obtain for the space components
\beq
\frac{\rmd v^i}{\rmd t}=\frac{\partial \Phi}{\partial x^{i}}-2({\pmb \beta}\times {\mathbf B})_i+2\frac{\partial A_{i}}{c\partial t}-\beta^i \frac{\partial \left(2\Psi+\Phi \right)}{c\partial t}\ \label{eq:geonew}
\eeq
(see e.g. \citet{Costa:2012cw} and also \citet{bini2008gravitational}, where the case $\Phi=\Psi$ is considered).  
Then, if we define the gravitoelectromagnetic fields as
\beq
\mb B= \bm \nabla \wedge \mb A, \quad \mb E= -\bm \nabla \Phi-\frac{2}{c} \ppar{\mb A}{t}, \label{eq:defEtime}
\eeq
the above equation (\ref{eq:geonew}) becomes
 {\beq
\frac{\rmd v^i}{\rmd t}=-E^{i}-2({\pmb \beta}\times {\mathbf B})_i+\frac 2 c \frac{\partial A_{i}}{\partial t}-\beta^i \frac{\partial \left(2\Psi+\Phi \right)}{c\partial t}. \label{eq:lor001}
\eeq}
As a consequence,  it is not warranted that the geodesic equation takes a Lorentz-like form if the fields are not static, due to the presence of the last term in (\ref{eq:lor001}): in order to evaluate its impact, we need to compare it with the gravitomagnetic terms $\displaystyle 2({\pmb \beta}\times {\mathbf B})_i$ and $\displaystyle \frac 2 c \frac{\partial A_{i}}{\partial t}$.   {As discussed for instance by  \citet{thorne1985laws} and \citet{Costa:2021atq}, the gravitomagnetic field can be originated by the translation of a source and by its spin.} In particular, the order of magnitude of the gravitomagnetic field due to the translation of a source with mass $M$ moving with speed $v_{s}$,  {at distance $r$,} is 
\beq
\left| B_{\mathrm{trans}} \right| \simeq  \frac{M v_{s}}{cr^{2}}. \label{eq:Btrans}
\eeq
As for the gravitomagnetic field of a spinning source, with angular momentum $\mb S$, radius $R$ and peripheral speed $v_{rot}$, we have
\beq
\left| B_{\mathrm{spin}} \right| \simeq  \frac{S }{cr^{3}} \simeq \frac{Mv_{rot}R}{c r^{3}}. \label{eq:Bspin}
\eeq
Hence, we see that
\beq
\left| {\pmb \beta}\times {\mathbf B} \right|_{\mathrm{trans}} \simeq \frac{M v v_{s}}{c^{2}r^{2}}, \quad \left| {\pmb \beta}\times {\mathbf B} \right|_{\mathrm{spin}} \simeq \frac{M v v_{rot}R}{c^{2} r^{3}}. \label{eq:Btransspin1}
\eeq
For a source like the Earth, the spin contribution is much lower than the translational one,  {since typically $R\ll r$ and $v_{rot}\ll v_{s}$}.
We can do similar estimates for the time variation of the vector potential $\mb A$, and we obtain:
\beq
\left|\frac{\partial A_{i}}{c\partial t} \right|_{\mathrm{trans}} \simeq \frac{M  v_{s}^{2}}{c^{2}r^{2}}, \quad \left|\frac{\partial A_{i}}{c\partial t} \right|_{\mathrm{spin}} \simeq \frac{M v_{s} v_{rot}R}{c^{2} r^{3}}. \label{eq:Attransspin1}
\eeq
As for the last term in (\ref{eq:lor001}), we have
\beq
\left|  \beta^i \frac{\partial \left(2\Psi+\Phi \right)}{c\partial t} \right| \simeq \frac{Mv v_{s}}{c^{2}r^{2}}. \label{eq:phipsit}
\eeq
Accordingly, we see that the latter contribution is of the same order as of the translational contribution in (\ref{eq:Btransspin1}):   {it can be neglected if we assume that the source is at rest or, keeping the spin contribution, when $r v_{s}\ll v_{rot}R$. In addition, we see that even for a source at rest, in general the term $\displaystyle \frac 2 c \frac{\partial A_{i}}{\partial t}$ cannot be neglected.}\\

The interaction of test masses with the gravitational field can be studied using a variational principle $\delta \int { L} dt=0$ starting from  the Lagrangian ${ L}=-mcds/dt$ which, according to the Eq. (\ref{eq:weakfieldmetric11}), is given by

\begin{equation}
{ L} = - mc^{2} \left[ 1-{v^{2}\over c^{2}} - {2 \over c^{2}} \left( 1+
{v^{2}\over c^{2}} \right) \Phi + {4\over c^{3}} {v_{i}A^{i}}
\right]^{1/2}, \label{eq:lagrangian1}
\end{equation}
which, up to linear order in $\Phi$ and ${\mathbf A}$, and taking the lowest order terms ${\mathbf v} / c$ multiplying the gravitoelectromagnetic potentials, we obtain
\begin{equation}
{\cal L} = - mc^{2} \left( 1- {v^{2} \over c^{2}} \right)^{1/2} + m \Phi - {2 m \over c}  v_{i}A^{i}. \label{eq:lagrangian2}
\end{equation}
The term added to the free-particle Lagrangian, $\displaystyle m \Phi - {2 m \over c}  v_{i}A^{i}$ describes the interaction of the test particle with the field: again, we see that  the gravitomagnetic charge is  twice the gravitoelectric one. Furthermore, we see that the canonical momentum $\mathbf P=\partial L/\partial \mathbf v$ is given by $\displaystyle \mathbf P=m\mathbf v-\frac{2m}{c}\mathbf A$. \\

 {As we are going to show, the geodesic equation takes the form of a Lorentz-like equation  when we use Fermi coordinates.} The latter are defined starting from the world-line of an observer and they allow to show that what an observer measures depends both on the background field where she/he is moving and on her/his  motion. Fermi coordinates are important in the measurement process because they have a concrete meaning, since they are the coordinates an observer would naturally use to make space and time measurements in the vicinity of her/his world-line. This is particularly relevant when dealing with gravitational waves. They  are usually studied in the transverse traceless (TT)  gauge coordinates (see \citet{flanagan2005basics}) which do not have a physical meaning: an approach to the study of gravitational waves using Fermi coordinates is discussed by \citet{Ruggiero:2021qnu}. More in general, Fermi coordinates allow to define a gravitoelectromagnetic analogy in full GR, on the basis of the properties of the Riemann curvature tensor\cite{Mashhoon:2003ax,costa2008gravitoelectromagnetic,Mashhoon:1996wa,Ramos:2006ho,Costa:2012cw,chicone2002generalized}. In particular, using Fermi coordinates $(cT,X,Y,Z)$, for geodesic observers the spacetime metric can be written (see e.g. \citet{Ruggiero_2020} and references therein) in the form given by Eq. (\ref{eq:weakfieldmetric11}), with
 {\begin{eqnarray}
\Phi (T, { X^{i}})&=&-\frac{c^{2}}{2}R_{0i0j}(T )X^iX^j, \label{eq:defPhiG}\\
A^{}_{i}(T ,{X^{i}})&=&\frac{c^{2}}{3}R_{0jik}(T )X^jX^k, \label{eq:defAG}\\
\Psi_{ij} (T, {X^{i}}) & = & -\frac{2c^{2}}{3}R_{ikjl}(T)X^{k}X^{l}. \label{eq:defPsiG}
\end{eqnarray}}
Notice that $R_{\alpha \beta \gamma \delta}=R_{\alpha \beta \gamma \delta}(T)$ is the Riemann curvature tensor evaluated along the reference geodesic, where $X^{i}=0$ and it depends on $T$ only, which is the observer's proper time. Then, keeping only terms to first order in $X^{i}$, we obtain the following expression for the gravitoelectromagnetic fields
\beq
E^{}_i(T ,{X^{i}})=c^{2}R_{0i0j}(T) X^j, \label{eq:defEIEG}
\eeq
and 
\beq
B^{}_i(T ,{X^{i}})=-\frac{c^{2}}{2}\epsilon_{ijk}R^{jk}_{\;\;\;\; 0l}(T )X^l. \label{eq:defBIBG}
\eeq
In this case, the third term in Eq. (\ref{eq:lor001}) vanishes, as it is second order in $X^{i}$, and the geodesic equation takes the form 
\beq
m\frac{\rmd {\mathbf V}}{\rmd t}=-m{\mathbf E}-2m \frac{{\mathbf V}}{c}\times {\mathbf B}. \label{eq:lor20}
\eeq
Notice that, in this case, $\mb V$ is the relative velocity with respect to a test particle on the reference world-line.\\
As we discussed in our previous papers, \citet{Ruggiero_2020} and \citet{Ruggiero:2021qnu}, this approach can be applied to the study of the spacetime around a world-line of an observer in the field of a plane gravitational wave.  Fermi coordinates were first applied by \citet{biniortolan2017} to the study of a plane gravitational wave. In particular if we consider a plane gravitational wave solution propagating   along the $x$ axis with frequency $\omega$, the line element in TT coordinates is given by
\beq
ds^2= -c^{2}dt^2+dx^2 +(1-h^{+})dy^2 +(1+h^{+})dz^2 -2h_{\times} dy dz\,, \label{eq:TTmetrica}
\eeq
where
\beq
h^{+}=A^{+}\sin \left(\omega t-kx  \right), \quad h^{\times}=A^{\times}\cos \left(\omega t-kx\right). \label{eq:gwsol2}
\eeq
In the above formulae   $A^{+}, A^{\times}$ are the amplitude of the wave in the two polarization states, while $k$ is the wave number. Starting from these definitions, and taking into account the fact that  in weak field approximation (up to linear order in the flat spacetime perturbations $h_{\mu\nu}$)   the Riemann tensor is invariant with respect to  coordinate transformations, from the definition of the gravitoelectromagnetic fields (\ref{eq:defEIEG})-(\ref{eq:defBIBG}), 
we obtain the following expressions in Fermi coordinates
\begin{widetext}
\beq
E^{}_{X}  = 0, \quad E^{}_{Y}  = -\frac{\omega^{2}}{2}\left[A^{+} \sin \left(\omega T \right)Y+A^{\times} \cos \left(\omega T \right) Z \right], \quad E^{}_{Z}  = -\frac{\omega^{2}}{2}\left[A^{\times}\cT Y-A^{+}\sT Z \right], \label{eq:campoE}
\eeq
\normalsize
\small
\beq
B^{}_{X}  = 0, \quad B^{}_{Y}  = -\frac{\omega^{2}}{2}\left[-A^{\times} \cT Y+A^{+} \sT Z \right], \quad B^{}_{Z}  = -\frac{\omega^{2}}{2}\left[A^{+}\sT Y+A^{\times}\cT Z \right]. \label{eq:campoB}
\eeq
\end{widetext}
Notice that the above expressions and, in particular the gravitomagnetic field, are explicitly time-dependent.

\begin{figure}[h]
\begin{center}
\includegraphics[scale=0.30,draft=false]{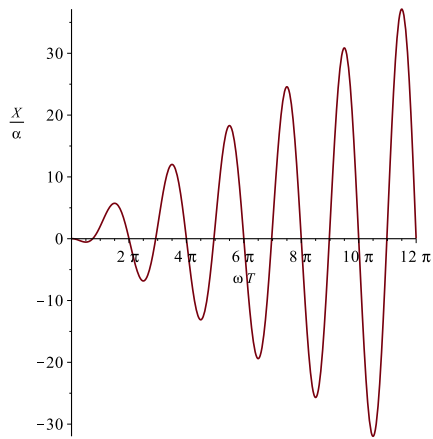}
\caption{The behavior of the particle coordinate parallel to the wave propagation direction: we set $\alpha=\frac{V^{2}_{0}}{c}A^{+}$. } \label{fig:oscillazione}
\end{center}
\end{figure}

Using this formalism, it is possible to describe  a new example of the action of the gravitomagnetic field of the wave on a moving  test mass, determined by the time-depending gravitomagnetic field.
We suppose that a particle is moving in the $YZ$ plane, hence orthogonally to the propagation direction of the wave.  Since the gravitomagnetic force is $\mb F^{B}=-2m\frac{\mb V}{c}\times {\mb B}$, the only component of this force is in the $X$ direction. To fix the ideas, let us suppose that, before the passage of the wave, the particle is moving with constant speed $V_{0}$, along the trajectory: 
\beq
X(T)=0,\quad Y(T)=0, \quad Z(T)=V_{0}T. \label{eq:motionlin}
\eeq
Also, we suppose that $A^{\times}=0$. Notice that we neglect the effects of the gravitoelectric field, which are confined to the $YZ$ plane. As a consequence, the only significant equation of motion turns out to be
\begin{eqnarray}
\dTT X &=& -\frac{V_{0}^{2}}{c} \omega^{2}A^{+} \left[\sT T \right], \label{eq:motionlinX} \\
\end{eqnarray}
Taking into account the initial conditions, we obtain the following solution:
\beq
X(T)=-\frac{V^{2}_{0}}{c}A^{+}\left[-\sT T+\frac{2}{\omega} \left(1-\cT \right)\right]. \label{eq:solmotionlin}
\eeq
We see that the passage of the wave provokes a motion of the particle out of the $YZ$ plane. The same qualitative result can be obtained for an arbitrary direction of the particle in the $YZ$ plane and, also, considering the  other polarization. A sketch of the motion induced by the wave is in Figure \ref{fig:oscillazione}. The oscillations have increasing amplitude, but they are physically limited since they are present only during the passage of the wave.  It is interesting to point out that the effects that are measured by current intereferometers are  in the $YZ$ plane, which is orthogonal to the wave propagation direction, since they are provoked by the gravitoelectric part of the wave field. On the other hand, this is effect (like other ones considered in \citet{Ruggiero_2020b,Ruggiero_2020}) is \textit{purely gravitomagnetic}: as we have seen, it is  simply described using this gravitoelectromagnetic approach, but it would be more complicated to understand in the framework of the TT gauge coordinates  that are usually employed to describe gravitational waves.

\section{Discussion and conclusions}\label{sec:disconc}

Many observational tests of General Relativity are performed in the so-called weak-field and slow-motion approximation: in other words, the gravitational field can be dealt with as a perturbation of flat spacetime and, moreover, both the sources and the test masses have slow speed compared to the speed of light. In this framework, Einstein equations and their solutions can be written in analogy with electromagnetism and a {linear} gravitoelectromagnetic formalism can be used. {In solving Einstein equations in this approximation, the Hilbert gauge condition is {often} used: we pointed out that, even if in the solutions for the metric tensor we neglect  terms that are $O(c^{-4})$, the gravitomagnetic potential and field  {are not necessarily stationary}. Different choices of the gauge conditions lead to a different form for the Maxwell-like equations for the gravitoelectromagnetic fields.}

{In addition, we considered a general  solution of Einstein equations that can be written in terms of a gravitelectric and gravitomagnetic potentials, 
and used the linear gravitoelectromagnetic analogy to study the motion of test masses. In particular, we discussed under which hypotheses the space components of the geodesic equation have a Lorentz-like form, and showed that this is possible when the sources of the gravitational field are at rest, or they are very slowly moving: if this is not the case, an extra non Maxwellian-like term is present. This is not surprising:  in fact, General Relativity and electromagnetism are obviously different  theories, and the fact that \textit{in given conditions} there is a similarity cannot be used to say that gravitation in the weak-field limit is completely analogous to electromagnetism. Moreover, we showed that we recover the Lorentz-like form for the geodesic equation in the framework of Fermi coordinates, to first order in the displacements from the reference world-line. As an application, we used this formalism  to study the motion of test masses in the field of a gravitational wave, and showed that, in doing so, purely gravitomagnetic effects arise that are more complicated to understand in the framework of transverse traceless coordinates, that are often used to study gravitational waves.}

We believe that what we have discussed in this  note can be useful both  to better understand the limitations of the gravitoelectromagnetic analogy and to exploit its capability to simplify the description of gravitational phenomena, in its range of applicability.

\bibliography{GEM_note}

\end{document}